\documentclass[useAMS,usenatbib,usegraphicx]{mn2e}

\usepackage{amssymb}
\usepackage{amsmath}
\usepackage{booktabs}
\usepackage{hyperref}
\usepackage{multirow}
\usepackage{times}

\title[X-ray time lags in PG\,1211+143]
  {X-ray time lags in PG\,1211+143}
\author[A.P. Lobban et al.]
  {A.P.~Lobban$^1$\thanks{e-mail: \href{mailto:a.p.lobban1@keele.ac.uk}{a.p.lobban1@keele.ac.uk}},
  S.~Vaughan$^2$, K.~Pounds$^2$, J.N.~Reeves$^{1,3}$ \\
  $^1$Astrophysics Group, School of Physical and Geographical Sciences, Keele University, Keele, Staffordshire, ST5 5BG, U.K. \\
  $^2$University of Leicester, X-Ray and Observational Astronomy Group, Department of Physics and Astronomy, Leicester, LE1 7RH, U.K. \\
  $^3$Center for Space Science and Technology, 1000 Hilltop Circle, University of Maryland Baltimore County, Baltimore, MD 21250, U.S.A.}
\date{\today; accepted for publication in MNRAS}

\pagerange{\pageref{firstpage}--\pageref{lastpage}} \pubyear{2016}

\def\LaTeX{L\kern-.36em\raise.3ex\hbox{a}\kern-.15em
    T\kern-.1667em\lower.7ex\hbox{E}\kern-.125emX}

\begin{document}

\label{firstpage}

\maketitle

\begin{abstract}

We investigate the X-ray time lags of a recent $\sim$630\,ks {\it XMM-Newton} observation of PG\,1211+143.  We find well-correlated variations across the {\it XMM-Newton} EPIC bandpass, with the first detection of a hard lag in this source with a mean time delay of up to $\sim$3\,ks at the lowest frequencies.  We find that the energy-dependence of the low-frequency hard lag scales approximately linearly with log($E$) when averaged over all orbits, consistent with the propagating fluctuations model.  However, we find that the low-frequency lag behaviour becomes more complex on timescales longer than a single orbit, suggestive of additional modes of variability.  We also detect a high-frequency soft lag at $\sim$10$^{-4}$\,Hz with the magnitude of the delay peaking at $\lesssim 0.8$\,ks, consistent with previous observations, which we discuss in terms of small-scale reverberation.

\end{abstract}

\begin{keywords}
 accretion, accretion discs -- galaxies, active galaxies: individual: PG\,1211+143, galaxies:Seyfert -- X-rays: galaxies
\end{keywords}

\section{Introduction} \label{sec:introduction}

Accretion on to black holes via an optically thick, geometrically thin accretion disc \citep{ShakuraSunyaev73} is thought to be the primary process through which luminous active galactic nuclei (AGN; $M_{\rm BH} \sim 10^{6-9}$\,$M_{\odot}$) and X-ray binaries (XRBs; $M_{\rm BH} \sim 10$\,$M_{\odot}$) are powered.  Both types of system are powerful sources of X-rays which are thought to originate via Compton-upscattering of thermal ultraviolet (UV) emission from the disc by a corona of hot electrons \citep{HaardtMaraschi93}.  The scattered X-rays then typically form an approximate power-law shape, which dominates the X-ray spectra of accreting black hole sources.  In addition to this, other routinely observed spectral features include a `soft excess' $< 2$\,keV \citep{ScottStewartMateos12}, a `Compton reflection' component $> 10$\,keV \citep{NandraPounds94} and various emission lines - in particular, strong emission from Fe\,K$\alpha$ fluorescence at $\sim$6.4\,keV \citep{GeorgeFabian91}.

Many details about the geometry and physics of the accretion flow are not well understood and disentangling such a wealth of emission components is challenging - however, X-ray timing provides an alternative diagnostic tool.  Recent observations of frequency-dependent X-ray time delays with {\it RXTE}, {\it XMM-Newton} and {\it NuSTAR} have revealed a range of different phenomena. Delays between variations in different X-ray energy bands -- with the hard X-ray variations lagging behind the correlated soft X-ray variations (`hard lags') -- are commonly detected at low frequencies ($\sim$10$^{-5}$--$10^{-4}$\,Hz) in bright, variable AGN (e.g. \citealt{PapadakisNandraKazanas01,VaughanFabianNandra03,McHardy07,LobbanAlstonVaughan14,Kara16,JinDoneWard17}).  Interestingly, the magnitude of the lag is often observed to increase with the separation of the energy bands.  Hard lags were first detected in XRBs (e.g. Cygnus X-1: \citealt{Cui97,Nowak99}) and it has been argued that the observed lags in AGN and XRBs are analogous but scaled to the appropriate time-scale depending on the size scales of the emitting region (e.g. \citealt{McHardy06}).

A number of models have been proposed to explain the delays, ranging from inverse-Compton scattering in the X-ray-producing corona (see \citealt{MiyamotoKitamoto89}) to X-ray reflection by the accretion disc \citep{KotovChurazovGilfanov01}.  The leading model to explain the observed low-frequency hard lags is the `propagating fluctuations' model whereby changes in the local mass accretion rate propagate inwards through the accretion disc, powering an extended corona of hot X-ray producing electrons \citep{Lyubarskii97}.  Here, stratification of the corona means that inward fluctuations firstly excite the outer, softer-X-ray-producing regions of the corona before driving emission with a harder spectrum from the inner regions, leading to an average hard delay.  This model successfully accounts for many observed variability properties of XRBs such as the energy-dependence of the power spectral density (PSD).

Further complexity is added to the lag behaviour by the detection of `soft lags', whereby more rapid soft X-ray variations lag behind the correlated harder X-ray variations \citep{DeMarco13}.  These soft lags are typically observed in the same AGN that exhibit hard lags but at higher frequencies.  They are often interpreted as a signature of the reverberation signal (see \citealt{Uttley14} for a review) as the primary X-ray emission is reflected by material close to the black hole (e.g. \citealt{ZoghbiUttleyFabian11,Fabian13}).  An alternative interpretation has been proposed by \citet{Miller10} whereby both the hard and soft lags arise from scattering of the primary X-ray continuum in more distant circumnuclear material tens to hundreds of gravitational radii from the central source.  However, a lot of support has recently built up behind the small-scale reverberation model through the discovery of Fe\,K features in lag-energy spectra (e.g. \citealt{AlstonDoneVaughan14,Kara14}).

In this paper we study the X-ray lags of the luminous narrow-line Seyfert galaxy / quasar, PG\,1211+143 ($z = 0.0809$; \citealt{Marziani96}).  PG\,1211+143 is X-ray bright with a typical X-ray luminosity of $\sim$10$^{44}$\,erg\,s$^{-1}$ and is also optically bright with a strong `Big Blue Bump'.  This source is well-known for its spectral complexity and is the archetypal source for displaying strong evidence for a highly-ionized, high-velocity outflow \citep{Pounds03, PoundsPage06, PoundsReeves07, PoundsReeves09}.  In addition, the source is observed to be relatively variable despite having a moderately large black hole mass of $M_{\rm BH} \sim 10^{7-8}$\,$M_{\odot}$ \citep{Kaspi00,Peterson04}.  \citet{DeMarco11} studied the X-ray time lags using {\it XMM-Newton} data from 2001, 2004 and 2007.  They discovered a lag at frequencies $\nu \lesssim 6 \times 10^{-4}$\,Hz, with the soft band (0.3--0.7\,keV) lagging behind the hard band (2--10\,keV) with a time delay of $\sim$500\,s.  However, those observations were relatively short in duration ($\sim$40--50\,ks) and spread apart over a number of years, only allowing frequencies down to $\nu \sim 10^{-4}$\,Hz to be accessed.  Here, we apply model-independent timing techniques to a new $\sim$630\,ks {\it XMM-Newton} campaign of PG\,1211+143.  These {\it XMM-Newton} data have previously been described in \citet{Lobban16a, Lobban16b} and \citet{Pounds16a, Pounds16b}.

\section{Observations and data reduction} \label{sec:observations_and_data_reduction}

PG\,1211+143 was observed seven times in 2014 with {\it XMM-Newton} (\citealt{Jansen01}) between 2014-06-02 and 2014-07-07.  Each observation had a typical duration of $\sim$100\,ks, except for the fifth observation (\textsc{rev}\,2664) which was shorter ($\sim$55\,ks), with a total duration of $\sim$630\,ks.  Here, we utilize data aquired with the European Photon Imaging Cameras (EPIC): the pn and the two Metal-Oxide Semiconductor (MOS) detectors, which were operated using the medium filter and also in large-window mode ($\sim$94.9 per cent `livetime' for the pn; $\sim$99.5 per cent for the MOS).  We processed all raw data using version 15.0 of the {\it XMM-Newton} Scientific Analysis Software (\textsc{sas}\footnote{\url{http://xmm.esac.esa.int/sas/}}) package, following standard procedures.  The details of the data and their reduction are provided in \citet{Lobban16a}\footnote{As pointed out in \citet{Zoghbi10}, the instrumental background rate can be high at energies $\gtrsim 8$\,keV, due to fluorescent Cu and Zn lines arising from the detector.  We utilize a large background region and note that the instrumental background region does not have a significant impact on our data.  See \citet{Pounds16a} for a comparison of source and background spectra.}.  In Table~\ref{tab:obs_log}, we provide an observation log of the {\it XMM-Newton} observations.

\begin{table}
\centering
\begin{tabular}{l c c c c}
\toprule
Date / & \multirow{2}{*}{EPIC} & \multirow{2}{*}{Total Duration} & \multirow{2}{*}{Count} & \multirow{3}{*}{Flux} \\
ObsID / & \multirow{2}{*}{Camera} & \multirow{2}{*}{[Net Exposure]} & \multirow{2}{*}{Rate} & \\
(Revolution) & & & & \\
\midrule
2014-06-02 & pn & 83 [77] & 3.97 & 1.10 \\
0745110101 & MOS\,1 & 78 [76] & 0.94 & 1.16 \\
(\textsc{rev}\,2652)  & MOS\,2 & 85 [83] & 0.93 & 1.17 \\
\midrule
2014-06-15 & pn & 100 [86] & 2.65 & 0.83 \\
0745110201 & MOS\,1 & 102 [97] & 0.57 & 0.83 \\
(\textsc{rev}\,2659)  & MOS\,2 & 102 [98] & 0.63 & 0.85 \\
\midrule
2014-06-19 & pn & 99 [90] & 3.31 & 0.94 \\
0745110301 & MOS\,1 & 101 [95] & 0.63 & 0.99 \\
(\textsc{rev}\,2661)  & MOS\,2 & 101 [95] & 0.76 & 1.01 \\
\midrule
2014-06-23 & pn & 96 [89] & 3.89 & 1.09 \\
0745110401 & MOS\,1 & 98 [95] & 0.82 & 1.11 \\
(\textsc{rev}\,2663)  & MOS\,2 & 98 [95] & 0.89 & 1.13 \\
\midrule
2014-06-25 & pn & 54 [51] & 5.01 & 1.36 \\
0745110501 & MOS\,1 & 56 [55] & 0.93 & 1.36 \\
(\textsc{rev}\,2664)  & MOS\,2 & 56 [55] & 1.16 & 1.44 \\
\midrule
2014-06-29 & pn & 92 [85] & 4.80 & 1.26 \\
0745110601 & MOS\,1 & 94 [91] & 0.98 & 1.31 \\
(\textsc{rev}\,2666)  & MOS\,2 & 94 [91] & 1.09 & 1.33 \\
\midrule
2014-07-07 & pn & 95 [89] & 3.73 & 1.02 \\
0745110701 & MOS\,1 & 97 [94] & 0.77 & 1.05 \\
(\textsc{rev}\,2670)  & MOS\,2 & 97 [94] & 0.85 & 1.08 \\
\bottomrule
\end{tabular}
\caption{A log of the {\it XMM-Newton} observations of PG\,1211+143 made in 2014 (see Section~\ref{sec:observations_and_data_reduction}).  The durations and net exposure times are given in ks, where `net exposure' refers to the integrated exposure time after accounting for the `dead time' of the detector.  All count rates and observed fluxes are calculated after filtering out background flares and are quoted over the full 0.2--10\,keV energy band in units of ct\,s$^{-1}$ and $\times 10^{-11}$\,erg\,cm$^{-2}$\,s$^{-1}$, respectively.  The observed fluxes are calculated from broad-band spectral modelling described in \citet{Lobban16a}.}
\label{tab:obs_log}
\end{table}

\section{Results} \label{sec:results}

\subsection{X-ray time lags}  \label{sec:x-ray_time_lags}

The lightcurve of PG\,1211+143 is shown in \citet{Lobban16a} (also later in Section~\ref{sec:x-ray_lightcurves}) and is found to be highly variable.  We followed the methods described in \citet{VaughanNowak97}, \citet{Nowak99}, \citet{Vaughan03}, \citet{Uttley11}, and \citet{EpitropakisPapadakis16}, which allows the variability in two distinct broad energy bands to be compared by calculating the cross-spectrum.  The seven lightcurves are split into a number of segments of identical length and, for each segment, their Discrete Fourier Transforms are computed.  These are then combined to form auto- and cross-periodograms and averaged over the number of segments.  This provides, as functions of Fourier frequency, estimates for the power spectra of the two bands, the coherence (see below) and time lags.

We utilize data from all three EPIC detectors (to maximize $S/N$), initially using 70\,ks segments, accessing frequencies down to $\nu \sim 1.4 \times 10^{-5}$\,Hz.  However, we must exclude data from the \textsc{rev}\,2664 observation, which is only $\sim$50\,ks in length.  We also exclude the first $\sim$5\,ks of \textsc{rev}\,2663 due to a large background flare\footnote{\textsc{rev}\,2661 suffers from multiple flares towards the end of observation.  However, this only results in $\lesssim 5$\,ks of flaring behaviour in our 70\,ks segment.  We include this observation in our timing analysis since it does not significantly affect the results.} (see fig.\,1 of \citealt{Lobban16a}).

In terms of assessing the reality of any measured time lags, one necessary criterion is that the light curves, at a given frequency, show some degree of coherence.  The coherence is calculated from the magnitude of the cross-periodogram \citep{VaughanNowak97} and is a measure of the linear correlation between two energy bands.  Its value should lie between 0 and 1, where a coherence of 0 means no correlation while a coherence of 1 signifies that the two energy bands perfectly linearly predict the variability in each other.  

\begin{figure}
\begin{center}
\hspace{-0.5em}
\rotatebox{0}{\includegraphics[width=8.4cm]{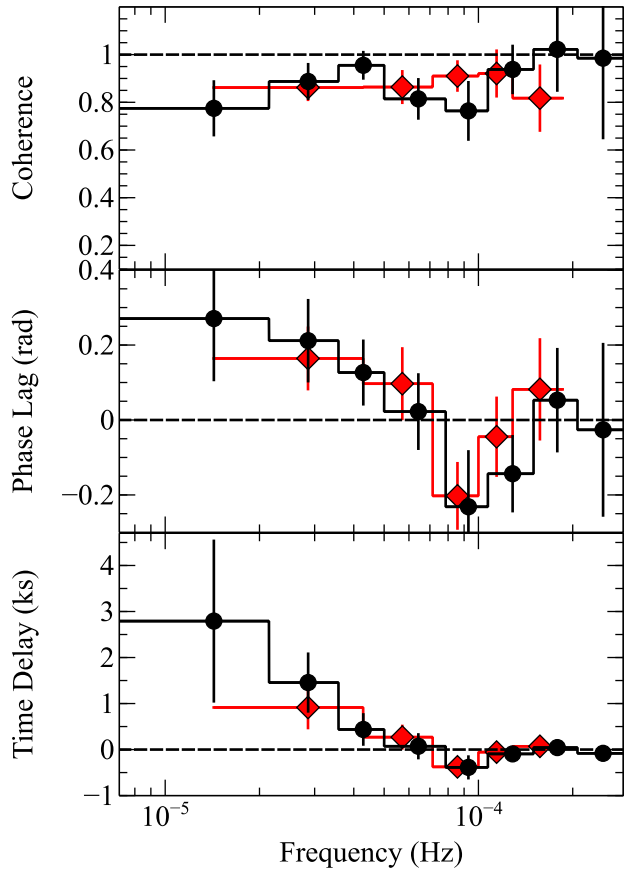}}
\end{center}
\vspace{-15pt}
\caption{The cross-spectral products for the soft (0.7--1.5\,keV) and hard (2--10\,keV) X-ray bands in PG\,1211+143 from EPIC-pn+MOS data averaged across all orbits.  Upper panel: the coherence between the two energy bands.  Middle panel: the phase lag between the two bands.  Lower panel: the time lag between the hard and soft band (a positive value denotes the hard band lagging behind the soft band).  The black circles and red diamonds represent data computed with 70\,ks segments and 35\,ks segments, respectively.}
\label{fig:lag-frequency}
\end{figure}

The magnitude and shape of the measured lag-frequency spectrum will, in general, be sensitive to the choice of energy bands.  We use two broad energy bands which provide a high level of coherence across a large range of frequencies: 0.7--1.5\,keV versus 2--10\,keV.  The mean frequency-dependence of the time lags from the 2014 observations is shown in Fig.~\ref{fig:lag-frequency}.  A more detailed energy dependence of the lags is presented in Section~\ref{sec:energy_dependence_lags}.  We used light curves extracted with $\Delta t = 100$\,s time bins and average the auto- and cross-periodograms over contiguous frequency bins, each spanning a factor of $\sim$1.4 in frequency.  The upper panel shows the coherence as a function of frequency, after Poisson noise correction \citep{VaughanNowak97}.  Note that the coherence is high ($\sim$0.8--1) at low frequencies and up to $\nu \sim 3 \times 10^{-4}$\,Hz, implying that the soft and hard bands correlate well on long time-scales (i.e. $\gtrsim 3$\,ks).  The coherence is not well constrained at higher frequencies, although we note that Poisson noise begins to dominate at $\nu \gtrsim 4 \times 10^{-4}$\,Hz, as shown by the power spectrum in \citet{Lobban16a}.

The middle and lower panels of Fig.~\ref{fig:lag-frequency} show the phase lags and time lags as a function of frequency, respectively.  While the two energy bands are consistent with having zero lag at the very highest frequencies (i.e. $\gtrsim 1.5 \times 10^{-4}$\,Hz), a significant hard lag\footnote{We use the convention that a `hard' (or `positive') lag signifies a delayed response of the hard energy band compared to the soft one.} is detected $\lesssim$7$\times 10^{-5}$\,Hz, which increases roughly as a power law to a maximum time delay of $\sim$3\,ks at $\sim$1.4$\times 10^{-5}$\,Hz.  This is the first detection of a low-frequency hard lag in PG\,1211+143.  Additionally, we observe a soft, negative lag at $\sim$9$\times 10^{-5}$\,Hz, with a time delay, $\tau = -410 \pm 220$\,s.  A soft lag in PG\,1211+143 was first detected by \citet{DeMarco11}.

At the lowest frequencies probed here (i.e. $\sim$1.4 $\times 10^{-5}$\,Hz; 70\,ks segments),  we are only afforded 1 Fourier frequency per $> 70$\,ks observation.  This results in only 6 raw frequencies contributing to the lowest-frequency bin.  We also computed the lags using 35\,ks segments which has the benefit of providing us with more segments to average over and allows us to include data from the shorter \textsc{rev}\,2664 observation.  The coherence, phase lags and time lags estimated from 35\,ks segments are superimposed on Fig.~\ref{fig:lag-frequency} in red and are largely consistent with those obtained with 70\,ks segments down to $\nu \sim 3 \times 10^{-5}$\,Hz.

\begin{figure}
\begin{center}
\hspace{-0.5em}
\rotatebox{0}{\includegraphics[width=8.4cm]{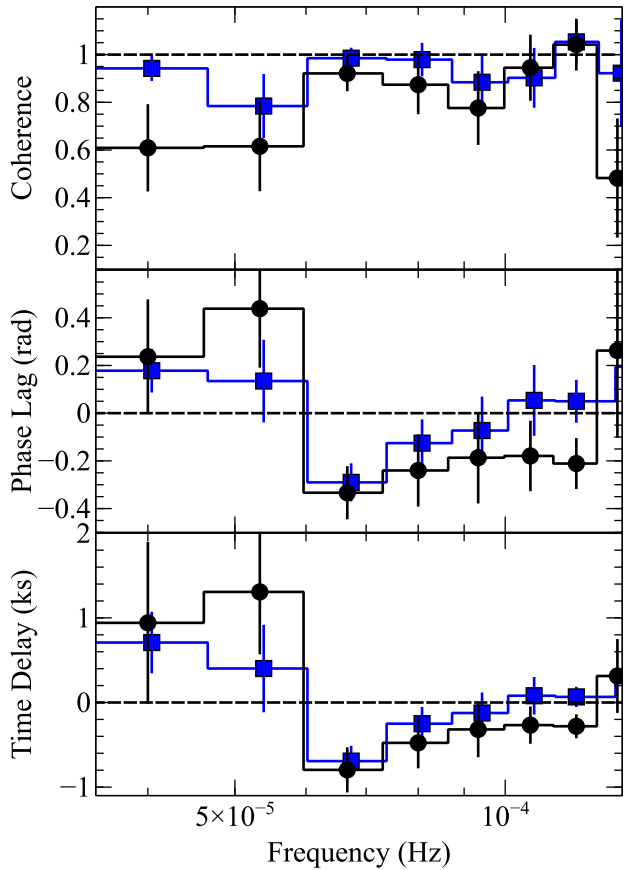}}
\end{center}
\vspace{-15pt}
\caption{A detailed view of the soft lag in PG\,1211+143, comparing the 0.2--0.7\,keV band with a harder 2--5\,keV band, estimated using 70\,ks segments (black circles).  Upper panel: the coherence between the two energy bands.  Middle panel: the phase lag between the two bands.  Lower panel: the time lag between the hard and soft band (a negative value denotes the soft band lagging behind the hard band).  The blue squares show the view of the soft lag between the 0.7--1.5 and 2--10\,keV bands for comparison with Fig.~\ref{fig:lag-frequency}.  A slight offset on the $x$-axis has been applied to the data to avoid overlapping error bars.}
\label{fig:soft-lag}
\end{figure}

In Fig.~\ref{fig:soft-lag}, we show a more detailed view of the soft lag, again with $\Delta t = 100$\,s time bins, but now with a finer frequency binning, with each bin spanning a factor of $\sim$1.1 in frequency.  This time we compare a soft 0.2--0.7\,keV band with a harder `continuum' band from 2--5\,keV.  This is for a more direct comparison with results from the literature, which we discuss in Section~\ref{sec:discussion_soft_lag}.  The lower panel of Fig.~\ref{fig:soft-lag} shows the time lag as a function of frequency where it can be seen that the soft lag extends over a wide range of frequencies ($\nu \sim 6 \times 10^{-5} - 1.5 \times 10^{-4}$\,Hz) with a peak time delay, $\tau = -790 \pm 260$\,s.  For comparison, we also overlay the view of the soft lag (blue squares) using the same energy bands as in Fig.~\ref{fig:lag-frequency} (i.e. 0.7--1.5 vs 2--10\,keV) with the same finer frequency binning.

\subsubsection{Energy dependence of the X-ray lags} \label{sec:energy_dependence_lags}

Motivated by the detection of a significant low-frequency hard lag in PG\,1211+143, we investigated the lag as a function of energy.  To calculate the lag-energy spectrum, a cross-spectral lag is calculated for a series of consecutive energy bands against a standard broad reference band over a given frequency range (e.g. \citealt{Uttley11}, \citealt{ZoghbiUttleyFabian11}, \citealt{AlstonDoneVaughan14}, \citealt{LobbanAlstonVaughan14}).  The choice of reference band sets the arbitrary lag offset in the resultant lag-energy spectrum.  Here, we generated cross-spectral products for 10 logarithmically-spaced energy bands from 0.2--10\,keV against a broad reference band consisting of the full 0.2--10\,keV energy range minus the energy band of interest\footnote{We also computed lag-energy spectra against a constant soft reference band of 0.2--0.7\,keV.  The shape of the lag-energy spectrum was consistent with that obtained with the broad reference band, just with an offset on the $y$-axis.}.  In this instance, a positive lag indicates that the given energy band lags behind the reference band.  We note that errors on the individual lag estimates in each band were calculated using the standard method (e.g. \citealt{BendatPiersol10})\footnote{The lightcurves involved in the lag estimate for each band are all highly correlated.  Since they are not independent realizations of a random process, we note that these errors are expected to be conservative since, between adjacent energy bins, they overestimate the scatter in the lags.}.

\begin{figure}
\begin{center}
\hspace{-0.5em}
\rotatebox{0}{\includegraphics[width=8.4cm]{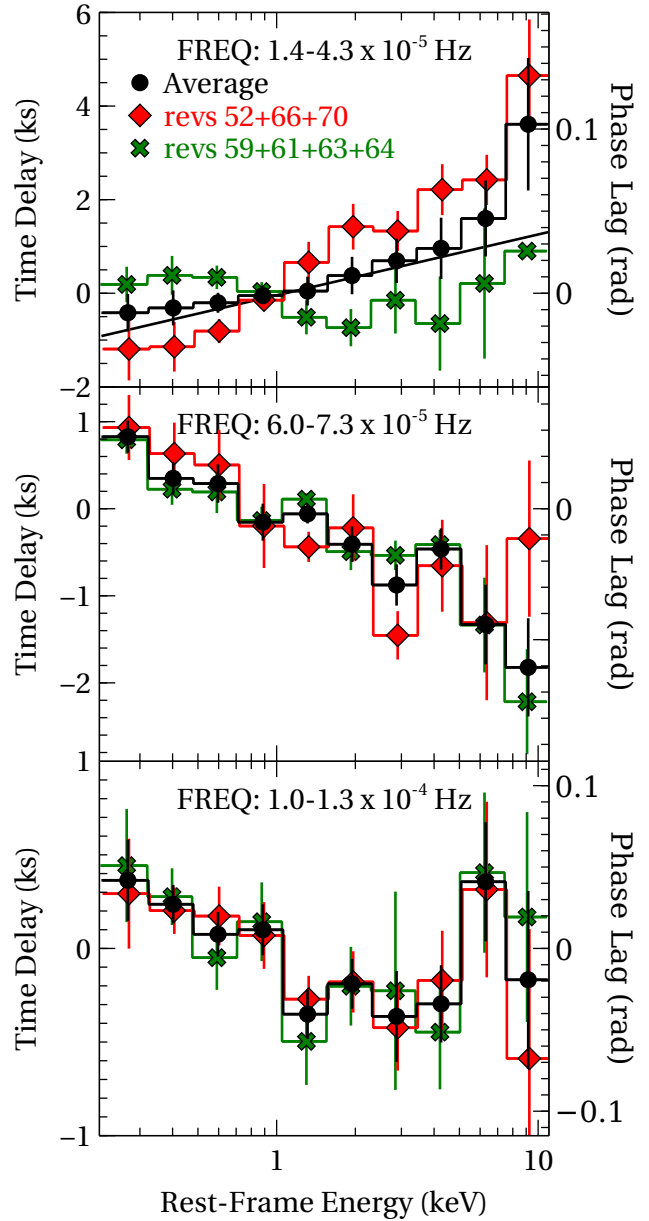}}
\end{center}
\vspace{-15pt}
\caption{The energy dependence of the lags in PG\,1211+143 against a broad reference band minus the band of interest.  The upper panel shows the lowest-frequency range where the hard lag is prominent ($\nu \sim 1.4 - 4.3 \times 10^{-5}$\,Hz).  The middle and lower panels cover the frequency range where the high-frequency soft lag dominates: $\nu \sim 6-7.3 \times 10^{-5}$\,Hz (where the soft lag is strongest) and $\nu \sim 1-1.3 \times 10^{-4}$\,Hz, respectively.  The black circles shows the lag-energy spectra averaged across all orbits while the red diamonds and green crosses show the lag-energy spectra averaged over \textsc{revs} 2652, 2666 and 2670 and \textsc{revs} 2659, 2661, 2663 and 2664, respectively.  The grey solid line shows a fit of the form: $\tau = A$log$(E) + B$ to the averaged data.}
\label{fig:lag-e}
\end{figure}

We computed lag-energy spectra over the lowest frequency bins obtained from our 35\,ks lag-frequency analysis in order to obtain better statistics (see Fig.~\ref{fig:lag-frequency}).  This covers the $\nu \sim 1.4$--$4.3 \times 10^{-5}$\,Hz frequency range and is shown in Fig.~\ref{fig:lag-e} (upper panel).  The mean low-frequency lag-energy spectrum averaged over all orbits is shown in black and suggests that the magnitude of the hard lag increases with the separation between energy bands, up to a time delay, $\tau$, of a few ks.

The shape of the time-averaged lag-energy spectrum suggests that the time lag scales approximately linearly with log($E$), as predicted for the propagating fluctuation model \citep{KotovChurazovGilfanov01}.  As such, we fitted the spectrum with a model of the form: $\tau = A$log$(E) + B$, where $A$ and $B$ are constants.  The fit is overlaid in Fig.~\ref{fig:lag-e}.  In these low-frequency data, we find that $A = 1.3 \pm 0.4$ and $B = 0.1 \pm 0.1$, with $\chi^{2}_{\nu} = 4.0/12$.  Similar behaviour can be observed in AGN such as Ark 564 \citep{Kara13} and IRAS 18325-5926 \citep{LobbanAlstonVaughan14} along with XRBs such as Cygnus X-1 \citep{Nowak99} and GX 339-4 \citep{Uttley11}.

\begin{figure}
\begin{center}
\hspace{-0.5em}
\rotatebox{0}{\includegraphics[width=8.4cm]{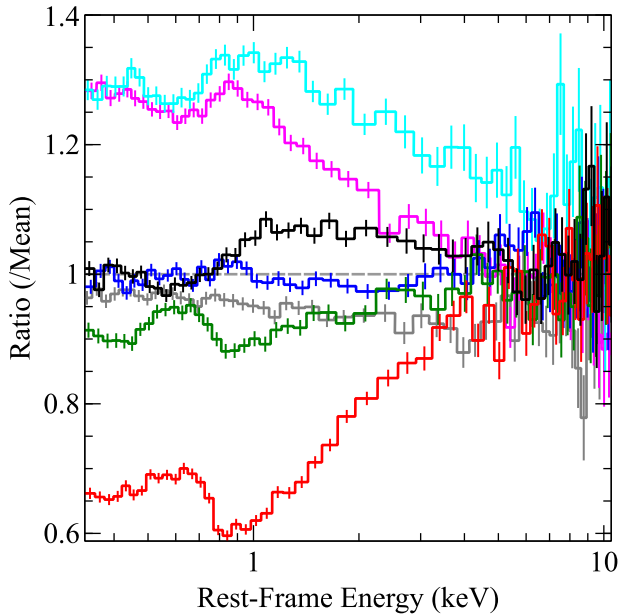}}
\end{center}
\vspace{-15pt}
\caption{The seven EPIC-pn spectra from 2014 (\textsc{rev}\,2652: black; \textsc{rev}\,2659: red; \textsc{rev}\,2661: green; \textsc{rev}\,2663: blue; \textsc{rev}\,2664: cyan; \textsc{rev}\,2666: magenta; \textsc{rev}\,2670: grey) plotted as a ratio against the time-averaged mean spectrum.}
\label{fig:ratio_spectrum}
\end{figure}

We searched for variations in the shape of the lag-energy spectrum by comparing the spectra obtained by combining only certain observations.  The source is observed to undergo a significant absorption event in \textsc{rev}\,2659, which, in particular, manifests itself as a deep absorption trough in the soft-band RGS spectrum [see \citealt{Lobban16a} and Reeves, Lobban \& Pounds (in press), a forthcoming paper detailing the inter-orbit spectral variability].  The absorption event lasts for the order of a $\sim$few days and, towards the end of the campaign, the spectrum returns to a state closely resembling the \textsc{rev}\,2652 spectrum at the start of the campaign.  All seven EPIC-pn spectra from 2014 are shown in \citet{Lobban16a} along with a high-flux / low-flux difference spectrum, which can be modelled with a steep power law ($\Gamma \sim 2.9$).  Another way to visualise the inter-orbit spectral differences is to use ratio spectra.  As such, we applied identical binning to all seven spectra.  The binning was relatively coarse to increase the $S$/$N$ and help visualise the effect.  We then divided each observation through by the time-averaged mean spectrum.  These ratio spectra are shown in Fig.~\ref{fig:ratio_spectrum}.

In terms of the low-frequency lags, a striking difference can be observed when comparing \textsc{revs} 2652, 2666 and 2670 with \textsc{revs} 2659, 2661, 2663 and 2664.  The lag-energy data are shown in Fig.~\ref{fig:lag-e} (upper panel).  In this instance, three orbits combine to produce the dominant steep shape of the spectrum.  The remaining orbits combine to produce a completely different low-frequency lag whose energy dependence is much flatter (relative to the broad reference band).    This behaviour is even more pronounced when probing the lower frequencies allowed with using 70\,ks segments with the hard lag reaching as high as $\sim$15--20\,ks in the \textsc{rev}\,2652+2666+2670 data when the energy-separation is largest - however, we stress that, in this case, there are only 3 `raw' measurements contributing to each bin and so our statistics are highly limited.

In the middle and lower panels of Fig.~\ref{fig:lag-e}, we show the energy dependence of the high-frequency soft lag over two separate frequency ranges.  The middle panel covers the $\nu \sim 6-7.3 \times 10^{-5}$\,Hz frequency range where the soft lag is strongest (see Fig.~\ref{fig:soft-lag}).  We again compute lag-energy spectra against a broad reference band consisting of the full 0.2--10\,keV band minus the band of interest\footnote{We also investigated the energy dependence of the soft lag against a constant hard reference band of 2--5\,keV.  The shape of the spectrum was unchanged although with a slight offset on the $y$-axis.}.  The soft lag smoothly increases in magnitude at lower energies, similar to the behaviour observed in sources such as Ark 564 \citep{Kara13}.  In the lower panel of Fig.~\ref{fig:lag-e}, we show the soft lag over a higher $\nu \sim 1-1.3 \times 10^{-4}$\,Hz frequency range.  While the shape of the spectrum is similar, there is a hint of a peak in the lag-energy spectrum at $\sim$6\,keV which appears, in shape, similar to the Fe\,K lags reported in other sources (e.g. \citealt{AlstonDoneVaughan14,Kara14}).  In both cases, we also show the energy dependence of the soft lag having combined observations consistently with the hard lag analysis.  We find that the energy-dependence of the high-frequency soft lag does not vary with flux or over time in these data, regardless of the choice of reference band.

\begin{figure}
\begin{center}
\hspace{-0.5em}
\rotatebox{0}{\includegraphics[width=8.4cm]{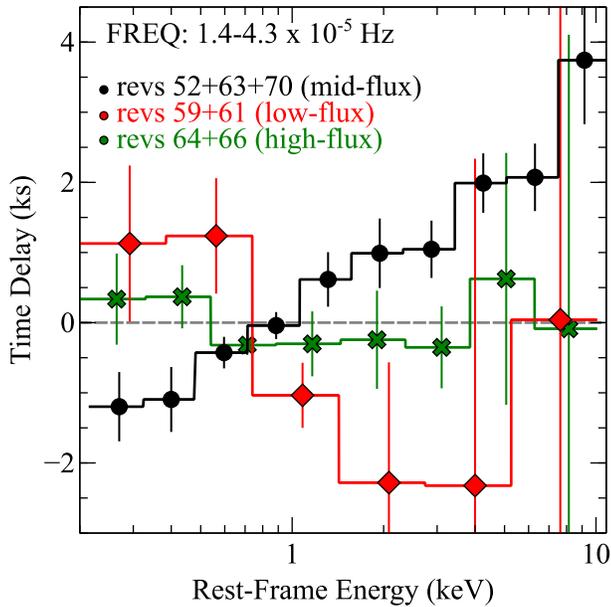}}
\end{center}
\vspace{-15pt}
\caption{The energy dependence of the low-frequency lags ($\nu \sim 1.4 - 4.3 \times 10^{-5}$\,Hz) in PG\,1211+143 against a broad reference band minus the band of interest.  The black circles show the medium-flux (`average') orbits while the red diamonds and green crosses show the lag-energy spectra computed from low-flux and high-flux orbits, respectively.}
\label{fig:lag-e_flux}
\end{figure}

While in Fig.~\ref{fig:lag-e}, we split the observations up according to an absorption event which occurs in \textsc{rev}\,2659, we also make an attempt to search for variations in the energy-dependence of the lag according to flux, using the same method described above.  As such, we combined orbits to compute low-flux (\textsc{revs} 2659+2661), medium-flux (\textsc{revs} 2652+2663+2670) and high-flux (\textsc{revs} 2664+2666) lag-energy spectra.  We note that \textsc{rev}\,2659 is markedly lower-flux and harder in spectral shape than any other orbit, but we cannot reliably measure the low-frequency lag using this observation alone as it would only offer a maximum of 2 `raw' frequency measurements, even using 35\,ks segments.  As such, we combine this observation with the next-lowest-flux orbit: \textsc{rev}\,2661, which also shows significant spectral signatures of the absorption event, which peaked a few days earlier.  The low-frequency lag-energy spectra are shown in Fig.~\ref{fig:lag-e_flux}, where the steep energy-dependence of the lag is dominated by the mid-flux observations.  The energy-dependence is much flatter in both the low-flux and, curiously, the high-flux cases, suggesting that the variable behaviour of the low-frequency hard lag may require a more complex explanation than simple linear variations with flux.  Finally, while we do not repeat the plots here, we also compared the energy-dependence of the soft lag at higher frequencies, finding no significant changes with flux or with the spectra shown in Fig.~\ref{fig:lag-e}.

\subsection{X-ray lightcurves} \label{sec:x-ray_lightcurves}

\begin{figure}
\begin{center}
\hspace{-0.5em}
\rotatebox{0}{\includegraphics[width=8.4cm]{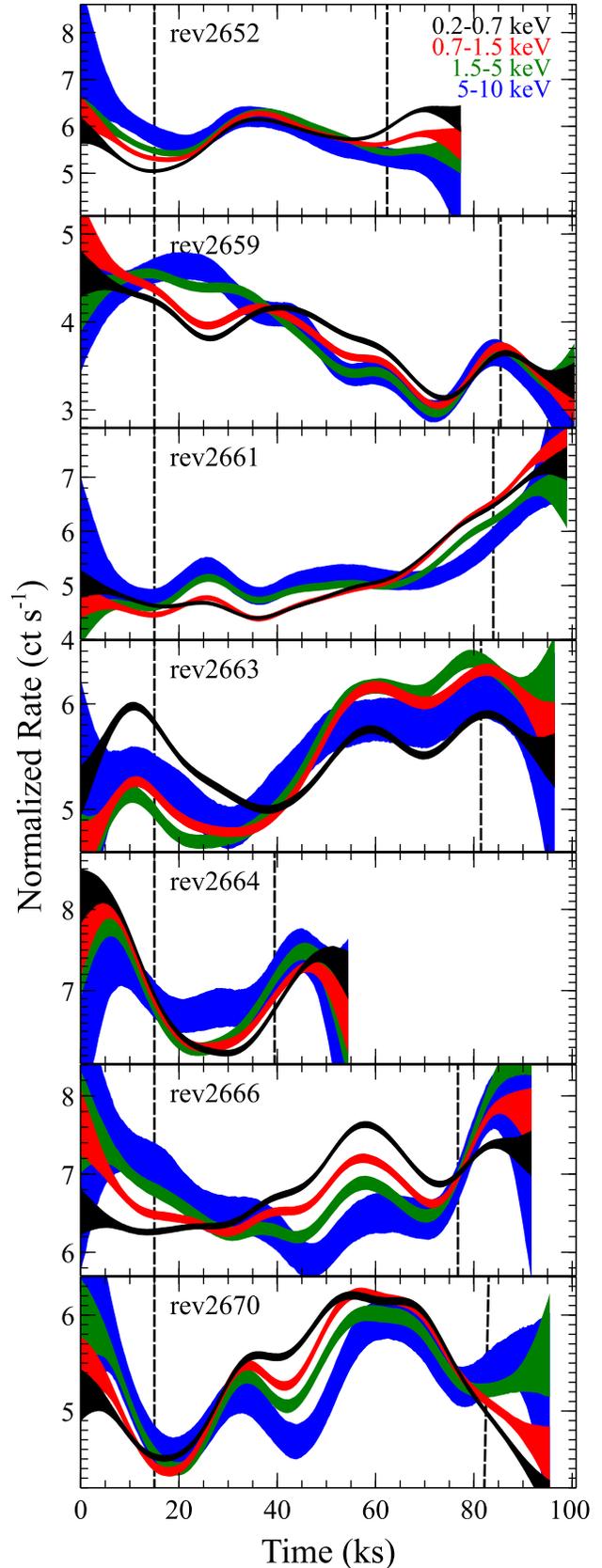}}
\end{center}
\vspace{-15pt}
\caption{The EPIC-pn+MOS lightcurves for each of the seven {\it XMM-Newton} observations from 2014 in four energy bands: 0.2--0.7 (black), 0.7--1.5 (red), 1.5--5 (green) and 5--10\,keV (blue).  Each lightcurve has been smoothed through a convolution with a Gaussian of width, $\sigma = 5$\,ks.  Within a given observation, the lightcurves were normalized to have the same mean count rate.}
\label{fig:lightcurve}
\end{figure}

The broad-band 0.2--10\,keV EPIC-pn lightcurve of PG\,1211+143 is shown in \citet{Lobban16a} with a 1\,ks timing resolution; the source varies by up to a factor of $\sim$2--3 over the course of the {\it XMM-Newton} campaign.  Here, part of our analysis involves investigating the properties of the lightcurve at low frequencies, where we detect significant time delays between energy bands.  Fig.~\ref{fig:lightcurve} shows the combined EPIC-pn+MOS lightcurves for each of the seven observations from 2014 across four different energy bands: 0.2--0.7, 0.7--1.5, 1.5--5 and 5--10\,keV.  Each lightcurve was extracted with a timing resolution of $dt = 100$\,s and has been convolved with a broad Gaussian of width, $\sigma = 5$\,ks, such that the high-frequency variations are smoothed out.  As such, only the low-frequency, longer-timescale variations remain.  Each lightcurve has been normalized such that its mean count rate is equal to that of the broad-band 0.2--10\,keV EPIC-pn+MOS rate for a given observation.  The lowest energy bands are found to be more variable with the 0.2--0.7 and 0.7--1.5\,keV bands varying by a factor of $\sim$3 (peak-to-peak) across all observations.  Meanwhile, the 1.5--5 and 5--10\,keV bands roughly vary by factors of $\sim$2 and $\sim$0.5, respectively.

To help assess the variability, we estimated 90\,per cent confidence intervals for each curve.  For a given lightcurve, we simulated 1\,000 curves of identical length and with the same timing resolution ($dt = 100$\,s), where the total number of counts in each bin was derived from a Poisson distribution assuming a mean identical to the total number of counts in the original observed bin.  We then convolved each simulated lightcurve with a Gaussian of width $\sigma = 5$\,ks and determined the 90\,per cent confidence limit by extracting the 5 and 95\,per cent values from the distribution of simulated light curves for each bin.  We also include vertical dashed lines in Fig.~\ref{fig:lightcurve} to represent the points along each lightcurve where the convolution kernel reaches the edge of the curve.  The half-width of the kernel is $3 \times \sigma$ and, hence, the convolution begins to become unreliable $\lesssim 15$\,ks from the end of each curve.

In general, it can be observed that the lightcurve has a similar shape across all energy bands.  However, a closer look at Fig.~\ref{fig:lightcurve} reveals more complex behaviour within individual observations.  For example, the first minimum in \textsc{rev}\,2652 appears to show the softer bands leading harder bands with progressively longer delays, with the three harder bands lagging behind the 0.2--0.7\,keV band by $\sim$3.4, $\sim$4.9 and $\sim$7.4\,ks, respectively.  However, roughly opposite behaviour can be observed in \textsc{rev}\,2663, where the softest band appears to lag behind the harder bands with a large delay of up to $\gtrsim 10$\,ks during the first minimum.  Curiously, these lags are then much less apparent throughout the rest of that particular observation - in particular, at the first maximum where there all four bands peak within $\sim$0.5\,ks of one another.  Further complex behaviour can also be observed in \textsc{rev}\,2670 where the softer bands lead the harder bands by $\sim$2\,ks during the first two minima, but with no obvious delay at the first maximum (after $\sim$35\,ks).

\begin{figure}
\begin{center}
\hspace{-0.5em}
\rotatebox{0}{\includegraphics[width=8.4cm]{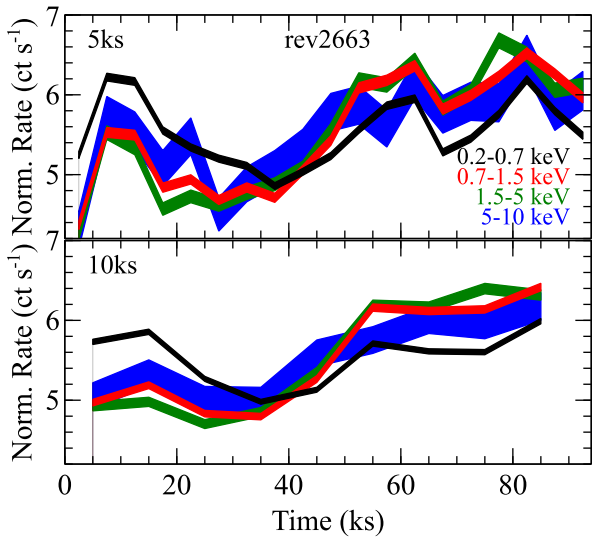}}
\end{center}
\vspace{-15pt}
\caption{The EPIC-pn+MOS lightcurves for \textsc{rev}\,2663 in four energy bands: 0.2--0.7 (black), 0.7--1.5 (red), 1.5--5 (green) and 5--10\,keV (blue).  The lightcurves are heavily binned with $dt = 5$\,ks (upper panel) and $dt = 10$\,ks (lower panel).  The lightcurves were normalized to have the same mean count rate.}
\label{fig:rev2663_5ks_10ks_lc}
\end{figure}

Gaussian smoothing is one of a variety of possible ways of filtering out high-frequency variations.  While having the advantage of producing a smooth output, we do also provide an alternative approach by computing heavily-binned lightcurves in the same four energy bands.  Here, we just focus on \textsc{rev}\,2663, which is one of the more interesting orbits, showing apparent variations in the behaviour at maxima and minima.  We show the lightcurves in 5\,ks and 10\,ks bins in Fig.~\ref{fig:rev2663_5ks_10ks_lc}, which demonstrate roughly the same effect we observe in Fig.~\ref{fig:lightcurve}.  In particular, the large soft delay occurring at the first minimum is clearly seen, along with the apparent lack of this behaviour during the large maximum $\sim$30\,ks later.  It is clear that, throughout the course of the campaign, not all maxima and minima behave in the same way.  As such, it is apparent that complex energy-dependent variations are taking place on long timescales in PG\,1211+143 with no particular low-frequency time lag persisting over time.

A natural question to arise given a) the complex behaviour of the lightcurves in different energy bands, and b) the apparent changes in the lag behaviour between orbits is whether there are any clear spectral variations that may offer clues or an explanation.  While all seven EPIC-pn spectra from 2014 and associated high-flux / low-flux difference spectra are shown in \citet{Lobban16a}, Fig.~\ref{fig:ratio_spectrum} shows all seven spectra as a ratio to the time-averaged mean.  It is interesting to note that the \textsc{rev}\,2659 spectrum (red) is noticeably harder than the others.  This is due to the spectrum being considerably more absorbed, which is evidenced by the sharp bite taken out of the spectrum at $\sim$0.8\,keV, predominantly arising from enhanced absorption from the unresolved transition array (UTA) from M-shell Fe transitions.  This can clearly be seen in figure 4 of \citet{Lobban16a} where a deep absorption trough can be observed in the \textsc{rev}\,2659 Reflection Grating Spectrometer (RGS; \citealt{denHerder01}) data.\footnote{Meanwhile, a detailed spectral study of the variability of the outflow on inter-orbital timescales will be presented in Reeves, Lobban \& Pounds (in press).}  While occurring sometime around \textsc{rev}\,2659, the strong absorption trough is still present, although weaker, in \textsc{rev}\,2661 and \textsc{rev}\,2663 and is coincident with the emergence of a slower moving, lower-ionization counterpart of the outflow (see \citealt{Pounds16a, Pounds16b}), while the soft excess is also diminished in flux.  A near-simultaneous dip in flux can also be observed in the long-term {\it Swift} \citep{Gehrels04} lightcurves presented in \citet{Lobban16a} (see figure 12) with the UV and X-ray fluxes dropping by $\sim$20 and $\sim$50\,per cent, respectively, roughly a day or two before \textsc{rev}\,2659.  Curiously, the enhanced absorption and emergence of the lower-ionization counterpart of the outflow appears to be simultaneous with the changing behaviour of the low-frequency hard lag, which becomes much weaker and has a flatter dependence on energy (see Fig.~\ref{fig:lag-e}: upper panel), before returning to its initial state towards the end of the campaign.

Finally, an additional test was performed to search for any obvious short-term inter-orbit spectral variability.  This was done by time-slicing all seven spectra into slices 10\,ks in length.  The shape of each spectral slice can generally be modelled with two simple power laws - a harder power law ($\Gamma \sim 1.7$) dominating $\gtrsim 1.5$\,keV and a soft power law ($\Gamma \sim 2.9$) dominating at low energies $\lesssim 1.5$\,keV.  However, the spectral variations on $\sim$10\,ks timescales are subtle and can be accounted for with small changes in the fluxes / photon indices of the two power laws (i.e. typically varying within $\sim$10\,per cent), which roughly track the behaviour of the soft and hard lightcurves.  As such, there is nothing obvious about the spectral shape on short timescales that could contribute to the changing lag behaviour while the individual within-observation slices do not have the $S$/$N$ to detect significant changes in discrete features (e.g. parameters of the high-velocity outflow).

\section{Discussion} \label{sec:discussion}

In this paper, we have presented an analysis of the short-timescale X-ray variability of PG\,1211+143 with {\it XMM-Newton}.  Through Fourier-based analysis, we have detected time lags in both the low- and high-frequency domains, which we now discuss in turn.

\subsection{Low-frequency time delays and the nature of the `average' lag} \label{sec:discussion_hard_lag}

In Section~\ref{sec:x-ray_time_lags}, we detected a hard lag where the 2--10\,keV band, on average, lags behind the softer 0.7--1.5\,keV band by $\sim$3\,ks at the lowest frequencies.  This is the first detection of a hard lag in this source.  Low-frequency hard lags are an important phenomenon as they may be ubiquitous in accreting black hole systems ranging from XRBs through to AGN with the frequency and amplitude of the lag scaling with black hole mass.  As such, they likely carry important information about the structure of accretion flows surrounding black holes.  

The leading model to explain the existence of low-frequency hard lags is the `propagating fluctuations' model \citep{Lyubarskii97}.  Such a model also predicts that the emitted X-rays follow a linear rms-flux relation (e.g. \citealt{ArevaloUttley06}), the likes of which has been observed in PG\,1211+143 (see \citealt{Lobban16a}).  Additionally, such a model is consistent with the observed energy dependence of the hard lag in the sense that the time delay appears to increase with the separation of the energy bands.  Given the similarities in the properties of the observed hard lags in XRBs (e.g. Cygnus X-1; \citealt{KotovChurazovGilfanov01}) and many variable AGN (e.g. \citealt{McHardy04}; \citealt{Fabian13}; \citealt{LobbanAlstonVaughan14}), it is conceivable that a similar mechanism is at work in all accreting black hole systems.  The observed time-averaged low-frequency time lags in PG\,1211+143 are consistent with such a model and such a detection is useful for helping populate the higher-luminosity / higher-$M_{\rm BH}$ end of the scaling relations.  

However, we show that more complex time-dependent behaviour of the low-frequency hard lag is also apparent.  By computing the low-frequency lag from combinations of specific observations, we find that the energy-dependence of the lag varies significantly, having a much steeper or flatter dependence (relative to the broad reference band) according to which orbits contribute to the lag (see Figs~\ref{fig:lag-e} and~\ref{fig:lag-e_flux}).  While this changing behaviour of the low-frequency lag is curious, we only have a limited sample size at the lowest frequencies and so exercise caution in its interpretation.  Nevertheless, variable lag behaviour could offer a unique insight into the accretion processes in AGN.  

One intriguing possible interpretation in the case of PG\,1211+143 may arise from the inter-orbit spectral behaviour which accompanies the lag variability.  As Fig.~\ref{fig:ratio_spectrum} helps illustrate (but also see \citealt{Lobban16a}), a significant absorption event occurred around \textsc{rev}\,2659, lasting until roughly \textsc{rev}\,2663.  Predominantly manifesting itself in the soft band through absorption of the UTA, the absorption event may hint at enhanced activity in the outflow, particularly as it is somewhat coincident with the emergence of a lower-velocity, lower-ionization counterpart of the outflow (see \citealt{Pounds16a, Pounds16b}; Reeves, Lobban \& Pounds in press).  This absorption event is accompanied by a simultaneous drop in flux in the UV and X-ray bands (by $\sim$20 and $\sim$50\,per cent, respectively; see the {\it Swift} lightcurves presented in \citealt{Lobban16a}) and a diminished flux in the soft excess.  Given that the low-frequency hard lag is not detected in \textsc{revs} 2659+2661 - and so is coincident with a significant absorption event - it is possible that the two events are linked, perhaps arising from a change / disruption in the inner accretion flow, before the hard lag re-emerges later in the campaign.  

In the context of highly variable absorption, it is possible that different physical processes are operating on different timescales - e.g. photoionization versus a longer recombination timescale.  An interesting case study is the narrow-line Seyfert galaxy, NGC\,4051, where \citet{AlstonVaughanUttley13} found the low-frequency lag to be variable and dependent on the source flux.  \citet{SilvaUttleyCostantini16} then investigated the same dataset using a detailed time-dependent photoionization code to predict the effects of intervening, ionized, absorbing material on the observed time lags.  Curiously, they found that a warm absorber with the same properties as those observed in NGC\,4051, can produce soft X-ray lags in the low-frequency domain, where the time delay arises from radiative recombination and/or photoionization as the gas varies in response to the ionizing continuum.  As such, the low-frequency lags may carry both the signature of a hard-lag-producing process intrinsic to the accretion flow and also a diluting soft lag associated with the warm absorber.  As such, it is possible that similar effects are in play in the case of PG\,1211+143, scaled according to the properties of the system and its outflow.

In order to interpret time lags physically, it is important to consider them in the time domain as well as in Fourier space.  As such, we created a series of Gaussian-smoothed lightcurves, shown in Fig.~\ref{fig:lightcurve} (see Section~\ref{sec:x-ray_lightcurves}).  The lightcurves show that the variability behaviour changes across maxima and minima from softer bands leading harder bands to harder bands leading softer bands to, occasionally, no clear lag at all.  This is also illustrated in Fig.~\ref{fig:rev2663_5ks_10ks_lc}, where a long, soft-band delay can be observed towards the beginning of \textsc{rev}\,2663, which is then no longer apparent during the latter half of the orbit.  So, while the low-frequency lags, when averaged across the entire 2014 campaign, appear consistent with the propagating fluctuations model, it is clear that more complex modes of behaviour may also be apparent on timescales of up to $\sim$days.  As such, an alternative interpretation to the inter-orbital lag variations described above is that we are observing a combination of low-frequency X-ray time lags, possibly operating out-of-sync or on different timescales or even entirely independently of one another.  Some support for this interpretation comes from Fig.~\ref{fig:soft-lag}, which shows a soft lead in the $\sim$4.5-6 $\times 10^{-5}$\,Hz band and a soft delay in the $\sim$6-7.2 $\times 10^{-5}$\,Hz band, both of which may be considered to be in the low-frequency domain. 

A natural question that may arise when one observes time delays is whether they are statistically significant or not.  In one sense of the word, one may ask whether the apparent variations are due to random fluctuations in the Poisson noise, when, in fact, the two lightcurves vary simultaneously.  Or, in other words, given infinite signal-to-noise, would the two lightcurves reach minima / maxima at the same time?  In the case of PG\,1211+143, it is clear that some of the variations in some given energy band do not coincide with the confidence bands of different energy bands (e.g. in \textsc{rev}\,2663; see Fig.~\ref{fig:rev2663_5ks_10ks_lc}.).  In a second sense of the term `significant', one could ask if the variations are due to intrinsic-but-random differences in the lightcurves, rather than a systematic delay - i.e. if given an infinitely-long lightcurve with an infinite number of maxima / minima, would one find the lags to be randomly distributed about zero?

It is important to note that when Fourier methods are employed in performing time lag analysis, the result is an estimate of an {\it average} lag, assuming stationary processes.  However, variability of AGN - and, indeed, XRBs - need not be so simple.  Given a non-linear or non-stationary response function between energy bands, it could be that minima exhibit delays in a different sense to maxima.  Then, given a sufficiently long dataset, one may find that the distributions of lags at minima and maxima were not centered on zero but also different.  However, then using Fourier methods to perform a standard lag analysis would yield some weighted average of the different lags intrinsic to the different physical processes.  As such, Fourier methods may have the potential to be misleading if the time lags are caused by anything more complicated than simple, linear response functions.  It may even be the case that there is no such thing as ``the lag''.  Indeed, if the observed lag is different, but nevertheless repeatable, at different parts of the lightcurve - for example, if maxima are different from minima and/or rises show different time delays to falls, then the average lag one estimates from Fourier analysis is some weighted average based on the shape of the lightcurve one happened to observe.

\subsection{The high-frequency soft lag} \label{sec:discussion_soft_lag}

We also detect the presence of a soft lag (i.e. softer X-rays lagging behind harder X-rays) at higher frequencies (see Figs~\ref{fig:lag-frequency} and~\ref{fig:soft-lag}).  The lag occurs at frequencies $\gtrsim 6 \times 10^{-5}$\,Hz with an averaged peak time delay, $\tau = -790 \pm 260$\,s (when comparing the 0.2--0.7 and 2--5\,keV bands).  Time lags in PG\,1211+143 were measured by \citet{DeMarco11} using {\it XMM-Newton} data acquired in 2001, 2004 and 2007.  While a soft lag was detected with a similar time delay, it occurred at frequencies $> 10^{-4}$\,Hz.  However, we note that the exact frequency and magnitude of the lag is sensitive to factors such as the chosen energy bands and frequency-binning.  Additionally, \citet{DeMarco11} noted some degree of variability of the soft lag between 2001, 2004 and 2007 and only had access to much shorter datasets (typically $\sim$40--50\,ks in length) and so were unable to access the much broader range of frequencies probed with our longer observations here.

\citet{DeMarco13} reported a scaling relation between the black hole mass and amplitude/frequency of the soft lag based on a sample of 15 AGN displaying high-frequency soft lags.  They found that the observed frequency, $\nu$, and time lag, $\tau$, are related to the black hole mass by the following relations: log\,$\nu = -3.50[\pm0.07] - 0.47[\pm0.09]$\,log\,$(M_{\rm BH})$ and log\,$|\tau| = 1.98[\pm0.08] + 0.59[\pm0.11]$\,log\,$(M_{\rm BH})$, where $M_{\rm BH}$ is the black hole mass in units of $10^{7}$\,$M_{\odot}$.  For the range of $M_{\rm BH}$ estimates for PG\,1211+143 ($\sim$10$^{7-8}$\,$M_{\odot}$), the scaling relations predict the frequency of the soft lag to lie in the range $\sim$8$\times10^{-5}$--$4 \times 10^{-4}$\,Hz, roughly consistent with what we observe.  However, this is more consistent with the higher end of the $M_{\rm BH}$ estimate range.  Meanwhile, the scaling relations predict the time delay to lie in the range $\sim$75--500\,s.  Again, this is roughly consistent with what we observe. However, the peak time delay of $\tau = -790$\,s between the 0.2--0.7 and 2--5\,keV bands (see Fig.~\ref{fig:soft-lag}) is more consistent with a higher black hole mass (e.g. $\sim$3$\times 10^{8}$\,$M_{\odot}$).  

An alternative hypothesis could be that the black hole mass in PG\,1211+143 is at the lower end of the estimate range (e.g. $\sim$10$^7$\,$M_{\odot}$) and that the magnitude of the soft lag is particularly large.  An additional way of estimating the black hole mass is to consider the well-established X-ray-rms--$M_{\rm BH}$ relation (e.g. \citealt{Ponti12}).  Here, a robust correlation is found between the rms variability in the X-ray band and the mass of the central black hole, with a higher-amplitude of variability found to be associated with lower-mass systems.  We test this in the case of PG\,1211+143 by calculating the normalized excess variance\footnote{The normalized excess variance is described in equation 1 of \citet{Ponti12} and is given by: $\sigma^{2}_{\rm rms} = 1/N\mu^{2} \sum^{N}_{i=1}[(X_{i} - \mu)^{2} - \sigma^{2}_{i}]$, where $N$ is the number of bins in the segment, $\mu$ is the mean count rate and $X_{i}$ is the count rate in a given bin with associated uncertainty, $\sigma_{i}$.  The uncertainty on each measurement of $\sigma^{2}_{\rm rms}$ is given by equation A.1 of \citet{Ponti12}, but also see \citet{Vaughan03}.} for a series of lightcurves with 250\,s binning in the 2--10\,keV energy band and with segment lengths of 20, 40 and 80\,ks, for a like-for-like comparison with \citet{Ponti12}.  We find measured values of $\sigma^{2}_{\rm rms}$ of $0.009 \pm 0.001$, $0.010 \pm 0.001$ and $0.015 \pm 0.001$ for the 20, 40 and 80\,ks-segment lightcurves, respectively.  In the 20\,ks case, the best-fitting relationship derived by \citet{Ponti12} from their sample of AGN is given by: log\,($\sigma^{2}_{\rm rms}$) $= (-2.13 \pm 0.14) + (-1.24 \pm 0.12)$ log($M_{\rm BH, 7}$).  The coefficients in the 40 and 80\,ks cases lie largely within these uncertainties.  Our measured rms values predict the mass of the black hole in PG\,1211+143 to be $0.88 \pm 0.24$, $1.02 \pm 0.23$ and $0.81 \pm 0.20 \times 10^{7}$\,$M_{\odot}$ for the three cases, respectively.  All of these lie at the lower end of the range of mass estimates, at odds with the measured properties of the soft lag.  So this could suggest that the observed X-ray variability in PG\,1211+143 is enhanced or the soft lag occurs at higher frequencies and with a larger magnitude than predicted.  Either way, we note that this may be one of the longest soft time delays detected in an AGN to date.

The most popular model to explain high-frequency soft lags involves reverberation of the primary X-ray emission by material close to the black hole, perhaps via reflection (e.g. \citealt{ZoghbiUttleyFabian11,Fabian13,Uttley14}).  In such a scenario, the observed time lag roughly corresponds to the distance between the primary and reprocessed emission sites.  In the case of PG\,1211+143, a time delay $\lesssim 1$\,ks would correspond to a distance of a few-to-tens of $r_{\rm g}$ for the given range of black hole mass estimates.  As such, the scaling of the characteristic timescales compared to lower-mass AGN (e.g. 1H0707-495; \citealt{ZoghbiUttleyFabian11}) may be consistent with the disc reverberation scenario.  Additionally, in a number of sources, the energy-dependence of the high-frequency soft lags have shown evidence for features in the Fe\,K band (e.g. see \citealt{AlstonDoneVaughan14,Kara14}), which have been associated with the small-scale reverberation model.  In Fig.~\ref{fig:lag-e}, we find there is a hint of a peak in the energy-dependence of the soft lag of PG\,1211+143 at $\sim$6\,keV, which appears similar to the Fe\,K lags detected in other sources.

So, while the behaviour of the high-frequency soft lag in PG\,1211+143 shares similarities with the small-scale reverberation model, we do note that ionized reflection does not appear to be the dominant component in the X-ray spectrum.  For example, the soft excess in the {\it XMM-Newton} spectrum is very smooth and does not appear to show any clear signatures of ionized reflection.  Indeed, \citet{Pounds16b} find that the soft excess is well-fitted with a smooth, steep power law, which we show in \citet{Lobban16a} to be the component primarily responsible for the inter-orbit spectral variability.  In \citet{Lobban16b} we fitted the broad-band spectrum with a relativistically-blurred ionized reflection model but find that we cannot simultaneously model the soft excess and the Fe\,K emission complex.  Having accounted for the now well-established components of the high-velocity outflow in the spectrum, we find that an additional component of ionized reflection is allowed by the data, but it is of only moderate strength and predominantly manifests itself in a component of excess emission just red-ward of the Fe\,K$\alpha$ complex (but do see Fig.~\ref{fig:lag-e} and the above discussion of Fe\,K lags).  So, while there may be a moderate component of ionized reflection in the broad-band X-ray spectrum of PG\,1211+143, it does not appear to be the dominant component.  Alternatively, it may be conceivable that the soft lag is still produced by material close to the black hole but instead via a secondary Comptonization component (e.g. as per \citealt{Done12}; \citealt{GardnerDone14}; \citealt{Rozanska15}), perhaps associated with the outer layers of the accretion disc or the surface of the inner regions of the outflowing wind.

\section*{Acknowledgements}

This research has made use of the NASA Astronomical Data System (ADS), the NASA Extragalactic Database (NED) and is based on observations obtained with {\it XMM-Newton}, an ESA science mission with instruments and contributions directly funded by ESA Member States and NASA.  This research is also based on observations with the NASA/UKSA/ASI mission {\it Swift}. APL acknowledges support from STFC consolidated grant ST/M001040/1 and JNR acknowledges financial support via NASA grant NNX15AF12G.  We wish to thank our anonymous referee for a thorough and constructive review of our paper.

\label{lastpage}


\begin{thebibliography}{}

\bibitem[\protect\citeauthoryear{Alston, Vaughan \& Uttley}{2013}]{AlstonVaughanUttley13}Alston W. N., Vaughan S., Uttley P., 2013, MNRAS, 435, 1511

\bibitem[\protect\citeauthoryear{Alston, Done \& Vaughan}{2014}]{AlstonDoneVaughan14}Alston W. N., Done C., Vaughan S., 2014, MNRAS, 439, 1548

\bibitem[\protect\citeauthoryear{Ar\'{e}valo \& Uttley}{2006}]{ArevaloUttley06}Ar\'{e}valo P., Uttley P., 2006, MNRAS, 367, 801

\bibitem[\protect\citeauthoryear{Bendat \& Piersol}{2010}]{BendatPiersol10}Bendat J. S., Piersol A. G., 2010, Random Data: Analysis and Measurement Procedures, 4th edn. Wiley, New York

\bibitem[\protect\citeauthoryear{Cui et al.}{1997}]{Cui97}Cui W., Zhang, S. N., Focke, W., Swank, J. H., 1997, ApJ, 484, 383

\bibitem[\protect\citeauthoryear{De Marco et al.}{2011}]{DeMarco11}De Marco B., Ponti G., Uttley P., Cappi M., Dadina M., Fabian A. C., Miniutti G., 2011, MNRAS, 417, 98

\bibitem[\protect\citeauthoryear{De Marco et al.}{2013}]{DeMarco13}De Marco B., Ponti G., Cappi M., Dadina M., Uttley P., Cackett E. M., Fabian A. C., Miniutti G., 2013, MNRAS, 431, 2441

\bibitem[\protect\citeauthoryear{den Herder et al.}{2001}]{denHerder01}den Herder J. W. et al., 2001, A\&A, 365, 7

\bibitem[\protect\citeauthoryear{Done et al.}{2012}]{Done12}Done C., Davis S. W., Jin C., Blaes O., Ward M., 2012, MNRAS, 420, 1848

\bibitem[\protect\citeauthoryear{Epitropakis \& Papadakis}{2016}]{EpitropakisPapadakis16}Epitropakis A., Papadakis I. E., 2016, A\&A, 591, 113

\bibitem[\protect\citeauthoryear{Fabian et al.}{2013}]{Fabian13}Fabian A. C., Kara E., Walton D. J., 2013, MNRAS, 429, 2917

\bibitem[\protect\citeauthoryear{Gardner \& Done}{2014}]{GardnerDone14}Gardner E., Done C., 2014, MNRAS, 442, 2456

\bibitem[\protect\citeauthoryear{Gehrels et al.}{2004}]{Gehrels04}Gehrels N., et al., 2004, ApJ, 611, 1005

\bibitem[\protect\citeauthoryear{George \& Fabian}{1991}]{GeorgeFabian91}George I., Fabian A. C., 1991, MNRAS, 249, 352

\bibitem[\protect\citeauthoryear{Haardt \& Maraschi}{1993}]{HaardtMaraschi93}Haardt F., Maraschi I., 1993, ApJ, 413, 507

\bibitem[\protect\citeauthoryear{Jansen et al.}{2001}]{Jansen01}Jansen F. et al., 2001, A\&A, 365, 1

\bibitem[\protect\citeauthoryear{Jin, Done \& Ward}{2017}]{JinDoneWard17}Jin C., Done C., Ward M., 2017, MNRAS, 468, 3663

\bibitem[\protect\citeauthoryear{Kara et al.}{2013}]{Kara13}Kara E., Fabian A. C., Cackett E. M., Uttley P., Wilkins D. R., Zoghbi A., 2013, MNRAS, 434, 1129

\bibitem[\protect\citeauthoryear{Kara et al.}{2014}]{Kara14}Kara E., Cackett E. M., Fabian A. C., Reynolds C., Uttley P., 2014, MNRAS, 439, L26

\bibitem[\protect\citeauthoryear{Kara et al.}{2016}]{Kara16}Kara E., Alston W. N., Fabian A. C., Cackett E. M., Uttley P., Reynolds C. S., Zoghbi A., 2016, MNRAS, 462, 511

\bibitem[\protect\citeauthoryear{Kaspi et al.}{2000}]{Kaspi00}Kaspi S., Smith P. S., Netzer H., Maoz D., Jannuzi B. T., Giveon U., 2000, ApJ, 533, 631

\bibitem[\protect\citeauthoryear{Kotov, Churazov \& Gilfanov}{2001}]{KotovChurazovGilfanov01}Kotov O., Churazov E., Gilfanov M., 2001, MNRAS, 327, 799

\bibitem[\protect\citeauthoryear{Lobban, Alston \& Vaughan}{2014}]{LobbanAlstonVaughan14}Lobban A. P., Alston W. N., Vaughan S., 2014, MNRAS, 445, 3229

\bibitem[\protect\citeauthoryear{Lobban et al.}{2016a}]{Lobban16a}Lobban A. P., Vaughan S., Pounds K., Reeves J. N., 2016, MNRAS, 457, 38

\bibitem[\protect\citeauthoryear{Lobban et al.}{2016b}]{Lobban16b}Lobban A. P., Pounds K., Vaughan S., Reeves J. N., 2016b, ApJ, 831, 201

\bibitem[\protect\citeauthoryear{Lyubarskii}{1997}]{Lyubarskii97}Lyubarskii Y. E., 1997, MNRAS, 292, 679

\bibitem[\protect\citeauthoryear{Marziani et al.}{1996}]{Marziani96}Marziani P., Sulentic J. W., Dultzin-Hacyan D., Calvani M., Moles M., 1996, ApJS, 104, 37

\bibitem[\protect\citeauthoryear{McHardy et al.}{2004}]{McHardy04}McHardy I. M., Papadakis I. E., Uttley P., Page M. J., Mason K. O., 2004, MNRAS, 348, 783

\bibitem[\protect\citeauthoryear{McHardy et al.}{2006}]{McHardy06}McHardy I. M., Koerding E., Knigge C., Uttley P., Fender R. P., 2006, Nature, 444, 730

\bibitem[\protect\citeauthoryear{McHardy et al.}{2007}]{McHardy07}McHardy I. M., Ar\'{e}valo P., Uttley P., Papadakis I. E., Summons D. P., Brinkmann W., Page M. J., 2007, MNRAS, 382, 985

\bibitem[\protect\citeauthoryear{Miller et al.}{2010}]{Miller10}Miller L., Turner T. J., Reeves J. N., Braito V., 2010, MNRAS, 408, 1928

\bibitem[\protect\citeauthoryear{Miyamoto \& Kitamoto}{1989}]{MiyamotoKitamoto89}Miyamoto S., Kitamoto S., 1989, Nature, 342, 773

\bibitem[\protect\citeauthoryear{Nandra \& Pounds}{1994}]{NandraPounds94}Nandra K., Pounds K. A., 1994, MNRAS, 268, 405

\bibitem[\protect\citeauthoryear{Nowak et al.}{1999}]{Nowak99}Nowak M. A., Vaughan B. A., Wilms J., Dove J. B., Begelman M. C., 1999, ApJ, 510, 874

\bibitem[\protect\citeauthoryear{Papadakis, Nandra \& Kazanas}{2001}]{PapadakisNandraKazanas01}Papadakis I. E., Nandra K., Kazanas D., 2001, ApJ, 554, 133

\bibitem[\protect\citeauthoryear{Ponti et al.}{2012}]{Ponti12}Ponti G., Papadakis I., Bianchi S., Guainazzi M., Matt G., Uttley P., Bonilla N. F., 2012, A\&A, 542, 83

\bibitem[\protect\citeauthoryear{Pounds et al.}{2003}]{Pounds03}Pounds K. A., Reeves J. N., King A. R., Page K. L., O'Brien P. T., Turner M. J. L., 2003, MNRAS, 345, 705

\bibitem[\protect\citeauthoryear{Pounds \& Page}{2006}]{PoundsPage06}Pounds K. A., Page K. L., 2006, MNRAS, 372, 1275

\bibitem[\protect\citeauthoryear{Pounds \& Reeves}{2007}]{PoundsReeves07}Pounds K. A., Reeves J. N., 2007, MNRAS, 374, 823

\bibitem[\protect\citeauthoryear{Pounds \& Reeves}{2009}]{PoundsReeves09}Pounds K. A., Reeves J. N., 2009, MNRAS, 397, 249

\bibitem[\protect\citeauthoryear{Pounds et al.}{2016a}]{Pounds16a}Pounds K., Lobban A., Reeves J., Vaughan S., 2016a, MNRAS, 457, 2951

\bibitem[\protect\citeauthoryear{Pounds et al.}{2016b}]{Pounds16b}Pounds K., Lobban A., Reeves J., Vaughan S., Costa M., 2016b, MNRAS, 459, 4389

\bibitem[\protect\citeauthoryear{Peterson et al.}{2004}]{Peterson04}Peterson B., Ferrarese L., Gilbert K. M., 2004, ApJ, 613, 682

\bibitem[\protect\citeauthoryear{R\'{o}\.{z}a\'{n}ska et al.}{2015}]{Rozanska15}R\'{o}\.{z}a\'{n}ska A., Malzac J., Belmont R., Czerny B., Petrucci P.-O., 2015, A\&A, 580, 77

\bibitem[\protect\citeauthoryear{Scott, Stewart \& Mateos}{2012}]{ScottStewartMateos12}Scott A., Stewart G., Mateos S., 2012, MNRAS, 412, 2633

\bibitem[\protect\citeauthoryear{Shakura \& Sunyaev}{1973}]{ShakuraSunyaev73}Shakura N. I., Sunyaev R. A., 1973, A\&A, 24, 337

\bibitem[\protect\citeauthoryear{Silva, Uttley \& Costantini}{2016}]{SilvaUttleyCostantini16}Silva C., Uttley P., Costantini E., 2016, A\&A, 596, 79

\bibitem[\protect\citeauthoryear{Uttley et al.}{2011}]{Uttley11}Uttley P., Wilkinson T., Cassatella P., Wilms J., Pottschmidt K., Hanke M., B\"{o}ck M., 2011, MNRAS, 414, 60

\bibitem[\protect\citeauthoryear{Uttley et al.}{2014}]{Uttley14}Uttley P., Cackett E. M., Fabian A. C., Kara E., Wilkins D. R., 2014, A\&ARv, 22, 72

\bibitem[\protect\citeauthoryear{Vaughan \& Nowak}{1997}]{VaughanNowak97}Vaughan B. A., Nowak M. A., 1997, ApJ, 474, 43

\bibitem[\protect\citeauthoryear{Vaughan, Fabian \& Nandra}{2003}]{VaughanFabianNandra03}Vaughan S., Fabian A. C., Nandra K., 2003, MNRAS, 339, 1237

\bibitem[\protect\citeauthoryear{Vaughan et al.}{2003}]{Vaughan03}Vaughan S., Edelson R., Warwick R. S., Uttley P., 2003, MNRAS, 345, 1271

\bibitem[\protect\citeauthoryear{Zoghbi et al.}{2010}]{Zoghbi10}Zoghbi A., Fabian A. C., Uttley P., Miniutti G., Gallo L. C., Reynolds C. S., Miller J. M., Ponti G., 2010, MNRAS, 401, 2419

\bibitem[\protect\citeauthoryear{Zoghbi, Uttley \& Fabian}{2011}]{ZoghbiUttleyFabian11}Zoghbi A., Uttley P., Fabian A. C., 2011, MNRAS, 412, 59

\end{thebibliography}
\end{document}